\documentclass[aps,prb,superscriptaddress,preprintnumbers,
showpacs,legalpaper,twoside,twocolumn,floatfix]{revtex4}

\usepackage{bm}
\usepackage{graphicx}
\usepackage{amsfonts}
\usepackage{amsmath}
\usepackage{amssymb}
\usepackage{amsthm}
\usepackage{bbm}
\usepackage{setspace}
\usepackage{helvet}
\newcommand{\bra}[1]{\langle #1 \bigr\rvert}
\newcommand{\ket}[1]{\bigl\lvert#1\rangle }
 
\newcommand{\kla}[1]{\left( #1 \right)}
\newcommand{\nif}[0]{\bigl\lbrace n_i^\text{f}\bigr\rbrace}

\theoremstyle{plain}

\begin{document}

\title{Dynamics of Localization Phenomena for Hardcore Bosons in Optical Lattices}

\author{Birger Horstmann}
\affiliation{Max-Planck-Institut f\"ur Quantenoptik, Hans-Kopfermann-Stra\ss e 1, 85748 Garching, Germany}
\affiliation{Institut f\"ur Theoretische Physik, Friedrich-Schiller-Universit\"at, Max-Wien-Platz 1, 07743 Jena, Germany}
\author{J. Ignacio Cirac}
\affiliation{Max-Planck-Institut f\"ur Quantenoptik, Hans-Kopfermann-Stra\ss e 1, 85748 Garching, Germany}
\author{Tommaso Roscilde}
\affiliation{Max-Planck-Institut f\"ur Quantenoptik, Hans-Kopfermann-Stra\ss e 1, 85748 Garching, Germany}


\begin{abstract}
 We investigate the behavior of ultracold 
bosons in optical lattices with a disorder potential generated via a secondary species 
frozen in random configurations. The statistics of disorder is 
associated with the physical state in which the secondary species
is prepared. The resulting
random potential, albeit displaying algebraic correlations, 
is found to lead to localization of all single-particle states. 
We then investigate the real-time dynamics of localization for a hardcore 
gas of mobile bosons which are brought into sudden interaction with 
the random potential. Regardless of 
their initial state and for any disorder strength, the mobile particles are found to reach a steady state 
characterized by exponentially decaying off-diagonal correlations 
and by the absence of quasi-condensation; when the mobile particles are 
initially confined in a tight trap and then released in the disorder 
potential, their expansion is stopped and the steady state is exponentially localized in real space, 
clearly revealing Anderson localization.  
\end{abstract}

\maketitle

\section{Introduction}

 The adiabatic loading of ultracold atoms and molecules in 
optical lattices represents a formidable opportunity to 
engineer strongly correlated states of quantum many-body 
systems with unprecedented control \cite{Jakschetal98, Greineretal02, Blochetal07, Lewensteinetal07}.
On the one hand, known fundamental models for the physics of solid 
state systems can be literally implemented in optical lattices 
\cite{toolbox, Janeetal03},
and the experimental detection of their ground state or thermal
equilibrium properties has the potential of responding to the 
lack of theoretical results due to fundamental limitations 
in the calculations, as those generally reported in 
fermionic systems or in frustrated quantum magnets.
On the other hand, optical lattice systems offer the 
further advantage of controlling the Hamiltonian parameters
in real time, and this enriches the range of 
correlated phases that can be implemented, if one can guide the 
evolution of the system towards an off-equilibrium 
state which is not necessarily an eigenstate of a known
Hamiltonian.  

 An intense activity has recently been focused on the experimental
implementation of disorder potentials in systems of trapped
ultracold bosons, both experimentally \cite{Lyeetal05, Clementetal05, Fortetal05, 
Schulteetal05, Schulteetal06,Fallanietal06} and theoretically \cite{Rothetal03,Rothetal03b,Damskietal03,Kuhnetal05,Sanchezetal07b,Clementetal06,Luganetal07,Sanchezetal06,Gimperleinetal05,Senguptaetal07,Sanperaetal04}. 
The introduction of tunable randomness in the system offers
the possibility of realizing Anderson localization of coherent
matter waves of weakly interacting bosons \cite{Clementetal05, Fortetal05, 
Schulteetal05} and the opportunity 
of investigating the interplay between strong localization 
and strong interaction, \emph{e.g.} in a deep optical lattice 
\cite{Fallanietal06}. The fundamental model describing this 
rich phenomenology in presence of a lattice is the Bose-Hubbard 
model in a random potential \cite{Fisheretal89}, where, beside
the conventional Mott insulating and superfluid phases, 
a Bose-glass phase appears, either associated with the 
fragmentation of weakly repulsive bosons into exponentially
localized states, or with the localization of collective
gapless modes for strong repulsion. 

 The disorder potential has so far been realized 
optically, either through laser speckles \cite{Lyeetal05, Clementetal05, 
Fortetal05, Schulteetal05} with or without an optical lattice,
or, in optical lattices, through a secondary incommensurate 
standing wave \cite{Fallanietal06}. In the case
of speckles, the typical length scale associated with the disorder potential
is quite extended with respect to the correlation 
length of the bosons, so that classical trapping rather than 
quantum localization is responsible for the observed suppression of
transport properties \cite{Schulteetal06,Sanchezetal07}. On the other hand,
the incommensurate superlattice of Ref.~\onlinecite{Fallanietal06}
realizes a potential which is strongly fluctuating over the distance
of a few lattice sites, but its \emph{pseudo}-disordered nature
requires to take also into account gapped insulating phases
at incommensurate fillings, competing with the Bose glass 
\cite{Reyetal06}. 

 An alternative proposal to create disorder in optical lattices 
involves the repulsive interaction with a secondary species of 
particles \cite{GavishC05, Paredesetal05}. 
If the two species correspond to two different hyperfine 
states of the atoms, the use of state-dependent optical lattices
\cite{Jakschetal99, Brennenetal99, Mandeletal03} allows one to first decouple the two species,
and then to selectively suppress the hopping of one of them, 
freezing it in a given quantum state $|\Phi^{\rm f}\rangle$. 

If the two species are subsequently brought into interaction, 
the mobile one evolves in a quantum superposition of all possible
realizations of the random potential associated with the
Fock components of  $|\Phi^{\rm f}\rangle$, and expectation values 
over the evolved state are therefore automatically averaged over
the disorder distribution \cite{Paredesetal05}. In this proposal
not only is the disorder potential strongly fluctuating
over the length scale of a few lattice sites, but in principle 
one can also vary its statistics by preparing the frozen bosons
in different states $|\Phi^{\rm f}\rangle$.
 
 In this paper we investigate the above proposal in detail
in the exactly solvable case of a one-dimensional gas 
of hardcore bosons on a lattice \cite{Liebetal61,Paredesetal04}. 
We consider both species of bosons (the mobile one and the frozen
one) to be hardcore repulsive, which is not only experimentally
feasible \cite{Paredesetal04, Weissetal04}, but it has three fundamental
theoretical advantages: 1) the state 
 $|\Phi^{\rm f}\rangle$ of the frozen bosons can be obtained
exactly from the Hamiltonian of the system before freezing,
and disorder averaging can therefore be accurately performed;
2) Jordan-Wigner diagonalization \cite{Liebetal61} allows to 
calculate the exact real-time evolution of the mobile particles 
\cite{Rigoletal04b,Rigoletal04,Rigoletal05} after they are brought 
into interaction with the frozen ones; 3) and, most importantly from a conceptual point of view, localization phenomena of hardcore bosons are perfectly understood in terms of Anderson localization of non-interacting fermions. While the observation of localization for interacting bosons is limited by screening of the disorder potential and the reduction of the healing length due to the interaction\cite{Lyeetal05, Clementetal05, Fortetal05, 
Schulteetal05, Schulteetal06,Fallanietal06}, many-body effects enter the system of non-interacting fermions only through Fermi statistics. 

In particular we focus on the case in which the 
frozen bosons are initially prepared in the superfluid
ground state  $|\Phi^{\rm f}\rangle$ of the hardcore-boson 
Hamiltonian at \emph{half} filling with periodic boundary conditions. 
The resulting disorder potential has the structure of a bimodal
random on-site energy, with algebraically
decaying correlations, but also with a very rich Fourier
spectrum dominated by short-wavelength components.
Most importantly, such a potential is found to lead to Anderson 
localization of all single-particle eigenstates (apart from 
possibly a set of zero measure) as expected for uncorrelated 
disorder \cite{Ishii72,Abrahamsetal79}. The evolution of 
the mobile bosons in such a potential is found to invariably lead 
to a disordered steady state with exponentially decaying
correlations and absence of quasi-condensation
for a wide variety of realistic initial conditions.
In particular phenomena of quasi-condensation 
in finite-momentum states \cite{Rigoletal04} and
fermionization \cite{Rigoletal05}, reported
upon expansion of the hardcore bosons from a 
Mott-insulating state and a superfluid 
state respectively, are completely washed out by
the disorder potential. Hence we can conclude that this setup 
allows for a robust implementation of a localized state 
as the off-equilibrium steady state of the system evolution. 

 This paper is structured as follows: Section \ref{System and Method}
describes the system of two bosonic species, the exact 
diagonalization method for the study of real-time evolution,
the Monte Carlo sampling of the disorder distribution, 
and the main features of the random potential;
Section \ref{Ground State Properties} is devoted to the
study of localization of the single-particle eigenstates 
in the random potential; Section \ref{Single Trap} investigates
the evolution of the mobile bosons after interaction with
the frozen ones when both species are prepared  
on a ring; finally, Section \ref{Transport Properties} 
is dedicated to the study of the expansion of the mobile
bosons in the random potential, starting from different
initial confined states.

\section{System and Method}
\label{System and Method}

In this section we present the system of two bosonic species which is used to 
study the effect of a disorder potential (subsection \ref{Hamiltonian Dynamics}). 
We then briefly describe the numerical procedure to exactly treat the 
equilibrium and out-of-equilibrium properties of the system
(subsection \ref{Jordan-Wigner Transformation and Implementation}), and 
the sampling of the disorder distribution (subsection \ref{Disorder Averaging}). 
Finally, we examine the nature of the correlations in the disorder potential 
created by a frozen species of hardcore bosons in subsection 
\ref{Characteristics of Disorder}.

\subsection{Hamiltonian Dynamics}
\label{Hamiltonian Dynamics}
The full system of two trapped interacting bosonic species in a one-dimensional
optical lattice is described by the Hamiltonian
\begin{equation}
\label{Hamiltonian}
{\cal H} = {\cal H}_0 + {\cal H}_{\rm f} + {\cal H}_{\rm int},
\end{equation}
where ${\cal H}_0$ is the Hamiltonian of the bosons which will
remain mobile,
\begin{eqnarray}
\label{H0}
{\cal H}_0  = &-& J\sum_{i=1}^L\left(a_i^\dagger a_{i+1}+{\rm h.c.}\right) \nonumber \\
&+& \sum_{i=1}^L \left[\frac{U}{2}n_i\bigl(n_i-1\bigr) + V \bigl(i-i_0\bigr)^2 n_i\right],
\end{eqnarray}
${\cal H}_{\rm f}$ is the Hamiltonian for the bosons to be frozen,
\begin{eqnarray}
\label{Hf}
{\cal H}_{\rm f} = &-& 
J^\text{f}\sum_{i=1}^L\left(a_i^{\text{f}\dagger} a_{i+1}^\text{f}+{\rm h.c.} \right)
\nonumber \\
&+& \sum_{i=1}^L\left[ \frac{U^\text{f}}{2}n_i^\text{f}\bigl(n_i^\text{f}-1\bigr) + 
V^\text{f} \bigl(i-i_0\bigr)^2 n_i^\text{f}\right],
\end{eqnarray}
and ${\cal H}_{\rm int}$ is the interaction Hamiltonian 
\begin{equation}
\label{Hint}
{\cal H}_{\rm int} = W\sum_{i=1}^L n_i n_i^\text{f}.
\end{equation}
Here $L$ is the number of sites of the system, $a_i^\dagger$ and $a_i$ are 
boson creation and annihilation operators and $n_i=a_i^\dagger a_i$ is the number 
operator for site $i$. Symbols with the superscript $\text{f}$ are the corresponding 
operators for the frozen bosons that create the
disorder potential. In Eqs.~(\ref{H0}),(\ref{Hf}) we also consider the
possibility of both species being confined by a parabolic trap 
 with different trapping strengths $V$,
$V^{\rm f}$. 

At times $t<0$ the two species of bosons are not interacting with each 
other ($W=0$), and they are prepared in the 
factorized ground state $\ket\Psi=\ket{\Phi^\text{f}}\otimes\ket\Phi$  of their
respective Hamiltonian with fixed numbers of particles $N$ and $N^\text{f}$. 
The ground state 
$\ket{\Phi^\text{f}}$ of the frozen particles can be decomposed in the Fock basis
\begin{equation}
\label{state frozen bosons}
\ket{\Phi^\text{f}}=\sum_{\{n_i^\text{f}\}}
c\kla{\left\{n_i^\text{f}\right\}}
\ket{\left\{n_i^\text{f}\right\}},
\end{equation}
where the sum extends over all Fock states, 
\begin{equation}
\nif=\left(n_1^\text{f},n_2^\text{f},\dots,n_L^\text{f}:
\sum_{i=1}^Ln_i^\text{f}=N^\text{f} \right).
\end{equation}

At some time $t\le 0$ the frozen bosons are made immobile ($J^\text{f}=0$), 
and subsequently at $t=0$ the interaction between the two species is turned 
on ($W>0$). 
Furthermore, releasing the mobile bosons from their trap ($V=0$) 
allows us to study their expansion properties.

The time evolution of the initially prepared state 
for $t>0$ is described by
\begin{equation}
\begin{split}
\label{time evolution parallel}
\ket{\Psi\kla{t}}&=e^{-i {\cal H} t} \sum_{\nif} c\kla\nif 
\ket\nif \otimes \ket{\Phi} \\
&=\sum_{\nif} c\kla{\nif,t} \ket\nif \otimes \ket{\Phi\kla{t,\nif}},
\end{split}
\end{equation}
where the coefficients $c(\nif,t) = e^{-i\phi(\nif)t} ~c\kla\nif$ 
have acquired a phase factor which will become irrelevant, and 
the state
\begin{equation}
\ket{\Phi\kla{t,\nif}}=
\exp\left\{-i \left[{\cal H}_0+{\cal H}_{int}\left(\nif\right)\right]t\right\}\ket{\Phi}
\end{equation}
represents the time evolution of the initial state of the mobile bosons 
interacting with a single Fock state $\ket\nif$ of the frozen 
bosons, which determines the static external potential.
We have used the property that, for $J^{\rm f}=0$,
$[{\cal H}_{\rm f},{\cal H}_0 + {\cal H}_{\rm int}]=0$.

Hence, equation \eqref{time evolution parallel} describes the parallel 
time evolution of the mobile bosons in a \emph{quantum superposition}
of different realizations of the disorder potential 
$V_i = V n_i^{\rm f}$, each appearing with a probability
$P\kla\nif = |c\kla\nif|^2 $. Remarkably, the time 
evolution of the expectation value of an operator 
$A$ acting only on the mobile bosons is automatically
averaged over the disorder statistics \cite{Paredesetal05}:
\begin{multline}
\label{expectation value}
\bra{\Psi\kla t} A\ket{\Psi\kla t} = \\ \sum_{\nif}\left|c\kla\nif\right|^2 
\bra{\Phi\kla{t,\nif}}A\ket{\Phi\kla{t,\nif}}.
\end{multline}

\subsection{Hardcore limit and Jordan-Wigner Transformation}
\label{Jordan-Wigner Transformation and Implementation}

From here onwards we will restrict ourselves to the exactly solvable
case of \emph{hardcore} bosons, obtained in the limit 
$U,U^{\rm f} \to \infty$ for filling smaller than 
or equal to one. It is convenient to incorporate the 
hardcore constraint directly in the operator algebra, 
passing to hardcore boson operators which \emph{anti}commute 
on the same site, $\{a_i,a_i^{\dagger}\}=1$, 
$\{a_i^{(\dagger)},a_i^{(\dagger)}\}=0$, and
$\{a_i^{\rm f},a_i^{{\rm f}\dagger}\}=1$,
$\{a_i^{{\rm f}(\dagger)},a_i^{{\rm f}(\dagger)}\}=0$.

Making use of the Jordan-Wigner 
transformation \cite{Liebetal61},
\begin{equation}
\label{Jordan-Wigner transformation}
a_i^\dagger=f_i^\dagger\prod_{k=1}^{i-1}e^{-i\pi f_k^\dagger f_k}, 
\hspace{0.5 cm} a_i=\prod_{k=1}^{i-1}e^{i\pi f_k^\dagger f_k}f_i, 
\end{equation}
one can map the hardcore bosons operators onto spinless ferm\-ion
operators 
$f_i^\dagger$ and $f_i$, obeying the same Hamiltonian as the
bosonic one, apart from a boundary term depending on the
number of particles in the case of periodic/antiperiodic
boundary conditions. The fermionic problem is exactly solvable,
and its eigenstates can be written in the following form
\begin{equation}
\label{matrix representation}
\ket{\Phi}=\prod_{m=1}^N\sum_{n=1}^L P_{nm}f_n^\dagger\ket 0,
\end{equation}
where the $N$ columns of the matrix 
$P_{nm} = \{{\bf P}\}_{nm}$ represent the first $N$ single-particle 
eigenstates. Following the recipe of Ref. \onlinecite{Rigoletal04b},
from the matrix $P_{nm}$ one can efficiently calculate
the one-particle density matrix (OPDM)
\begin{equation}
\label{OPDM}
\rho_{ij}=\langle a_i^\dagger a_j\rangle,
\end{equation} 
and hence the momentum distribution
\begin{equation}
\langle n_k\rangle =\frac{1}{L}\sum_{m,n=1}^Le^{-ik\kla{m-n}}
\langle a_m^\dagger a_n\rangle
\end{equation} 
which represents a fundamental observable in trapped 
atomic systems. At a more fundamental level,
the knowledge of the eigenvalues $\lambda_\eta$
of the OPDM, associated to eigenvectors $\phi_i^\eta$
also known as natural orbitals (NO), allows one
to rigorously study condensation phenomena
through the scaling of the maximum eigenvalue $\lambda_0$ 
with the particle number \cite{Penroseetal56,Rigoletal04b}. 
In absence of an external potential 
the OPDM of the hardcore bosons decays algebraically 
as $\rho_{ij}\sim {\left|i-j\right|}^{-\alpha}$, where $\alpha=0.5$,
signaling off-diagonal quasi-long-range order in the system \cite{Korepinetal93}. 
Correspondingly the occupation of the $k=0$ momentum state, which coincides with
the natural orbital with largest eigenvalue for a translationally
invariant system, scales as $n_{k=0}\sim \sqrt{N}$ with the particle number $N$
for any fixed density $n=N/L<1$, 
namely it exhibits \emph{quasi-condensation}. 
Remarkably, quasi-long-range order and quasi-condensation (in the form of a $\sqrt{N}$
scaling of the largest eigenvalue $\lambda_0$ of the OPDM) survive 
also in presence of a trapping potential $V \bigl(i-i_0\bigr)^a$ when the 
particle number is increased and correspondigly the trap
strength $V$ is decreased so that the characteristic density
in the trap
\begin{equation}
\label{particle density trap}
\tilde\rho=N \left(\frac{V}{J}\right)^{1/a},
\end{equation}
is kept constant and smaller than a critical value $\rho_c$
($\approx 2.6$ for $a=2$) to
avoid formation of a Mott plateau in the trap center 
\cite{Rigoletal04b}.

 Finally, the exact solution of the fermionic Hamiltonian
allows one to calculate the real-time evolution of the
fermionic wavefunction, and in particular of the
$P$ matrix as
\begin{equation}
\label{time evolution}
{\bf P}\kla{t}=e^{-i{\bf H}t}{\bf P}
\end{equation}
with the single-particle time evolution operator given by 
\begin{equation}
\label{time evolution operator}
\kla{e^{-i{\bf H}t}}_{ij}=\bra{0}f_ie^{-i{\cal H} t} f_j^\dagger\ket{0}.
\end{equation} 
Making use of this approach, Refs.~\onlinecite{Rigoletal04,Rigoletal05} 
have shown that quasi-condensation is a robust feature of the system 
after expansion starting from an initially trapped
quasi-condensed state, and it is even dynamically recovered 
when the initial state before expansion is a fully incoherent Mott 
insulator state.

\subsection{Disorder Averaging}
\label{Disorder Averaging} 

Eq.~(\ref{expectation value}) shows that, ideally, the unitary evolution of the system explores all possible realizations of the disordered potential at once. Numerical calculations
based on a matrix-product-state representation of the system
state also enjoy this feature of "quantum parallelism" of 
the Hamiltonian evolution \cite{Paredesetal05} by treating
the disorder potential as a quantum variable in the system. 
In this paper we use the more traditional approach of
exactly calculating the ground-state properties and the 
Hamiltonian evolution of hardcore bosons for a 
single realization of disorder at a time, averaging then
over the disorder distribution through Monte Carlo importance 
sampling. Accepting the overhead of disorder averaging, 
this approach has the advantage that, unlike the method
of Ref.~\onlinecite{Paredesetal05}, the time evolution is exact 
for arbitrarily long times.

 According to Eq.~\eqref{expectation value}, the weights of the disorder 
confi\-gurations are defined by the coefficients of the Fock-state
decomposition for the initial state of the 
frozen bosons through $P(\nif) = \bigl\lvert c\bigl(\nif\bigr)\bigr\rvert^2$. 
Introducing the bosonic Fock state
\begin{equation}
\label{Fock state}
\ket\nif=a_{i_1}^\dagger a_{i_2}^\dagger \dots a_{i_{N^{\rm f}}}^\dagger\ket{0}
\end{equation}
and the $N^{\rm f}\times L$ matrix 
$Q_{nm} =\{{\bf Q}\}_{nm}= \delta_{i_n,m}$, after some algebra we get 
\begin{equation}
\label{disorder determinant matrix}
c\bigl(\nif\bigr) = \langle \nif | \Phi_{\rm f} \rangle = 
\det \left({\bf Q}^{\dagger} {\bf P}\right).
\end{equation}

The energy eigenstates of noninteracting fermions with periodic boundary 
conditions, contained in the columns of the ${\bf P}$ matrix,
are given by plain waves. Thus it can be shown that the Slater determinant of 
Eq.~\eqref{disorder determinant matrix} is a Vandermonde determinant, 
which can be evaluated analytically. The disorder weights finally become
\begin{equation}
\label{cnif correlated}
\bigl\lvert c\kla{\nif}\bigr\rvert^2=\frac{1}{L^{N^{\rm f}}}
\prod_{1\le n < m\le N^{\rm f}}\sin^2\left[\frac{\pi}{L}\kla{i_n-i_m}\right].
\end{equation}

 Throughout the paper we will compare the effect of the disorder potential
 generated by the frozen superfluid state of $N^{\rm f}$ bosons with 
 that of the potential generated
 by $N^{\rm f}$ \emph{fully uncorrelated} frozen bosons \cite{remarkuncorrelated}, with 
 the flat distribution
 
  \begin{equation}
  \label{cnif uncorrelated}
\bigl\lvert c\kla{\nif}\bigr\rvert^2=
(N^{\rm f}!(L-N^{\rm f})!) / L!~.
\end{equation}

 Even when equipped with the exact statistics of disorder as in 
 Eqs.~\eqref{cnif correlated} and \eqref{cnif uncorrelated}, 
 it is generally hopeless to fully average the 
 results of any calculation on the $L!/(N^{\rm f}!(L-N^{\rm f})!)$
 disorder configurations. Hence we opt for Monte Carlo importance sampling, 
 where starting from an initially chosen random 
 configuration $\ket\nif$, we propose a new one $\ket{\nif'}$
 (e.g. by swapping the occupation of two sites)
 and accept it with Metropolis probability 
 $p = \min\left(\lvert c(\nif')\bigr\rvert^2/\lvert 
 c\kla{\nif}\bigr\rvert^2,1\right)$.
The real-space properties are typically averaged over $10^5$
disorder realizations, obtained one from the other
by updating $O(L)$ sites, while the OPDM is averaged
over $10^2$ realizations (due to the computational overhead of the
OPDM calculation). This provides full convergence of the
disorder-averaged quantities.

\subsection{Characteristics of Disorder}
\label{Characteristics of Disorder}

A fundamental figure of merit for the proposed 
setup to introduce disorder in optical lattices is 
represented by the correlation properties of the 
disorder potential generated by the frozen particles.  
As discussed in the introduction, the aspect of 
correlations represents one of the main weaknesses 
of the optical implementation of disorder through laser speckles 
\cite{Schulteetal06}. In particular, we focus on 
the density-density correlation function
\begin{equation}
\label{correlation function}
\begin{split}
C_{r}\kla{\ket{\Phi^f}}& = 
\frac{\frac{1}{L}\sum_{i=1}^L\left\langle\kla{n^{\rm f}_i-n^{\rm f}}
\kla{n^{\rm f}_{i+r}-n^{\rm f}}\right\rangle}{\frac{1}{L}\sum_{i=1}^L
\left\langle\kla{n^{\rm f}_i-n^{\rm f}}^2\right\rangle},
\end{split}
\end{equation}
where $n^{\rm f}=\langle n_i^{\rm f}\rangle=N^f/L$. This correlation function for the frozen hardcore bo\-sons 
is identical to the one for the corresponding fermions, and
can be calculated exactly in the homogeneous system ($V^\text{f}=0$) 
with periodic boundary conditions; for $N^{\rm f} < L$, namely
in the superfluid state of hardcore bosons,
\begin{equation}
\label{correlation function no superlattice}
C_r\kla{\ket{\Phi^f}}=\begin{cases}
1&\mbox{for } r=0\\
-\left[N^{\rm f}\bigl(L-N^{\rm f}\bigr)\right]^{-1} \frac{\sin^2\bigl(\pi\frac{N^{\rm f}r}{L}\bigr)}{\sin^2\bigl(\pi\frac{r}{L}\bigr)}&\mbox{for } r\ge 1\\
\end{cases}
\end{equation}
Hence, in the limit $r\ll L$ (which is always satisfied in the 
thermodynamic limit) the correlator decays algebraically like 
$C_r\kla{\ket{\Phi^f}} \sim r^{-2} $ with a superimposed
oscillation at twice the Fermi wavevector $k_{\rm F} = \pi n^{\rm f}$. The negative values 
of the correlator can be understood as resulting from
an effective long-range repulsion between the hardcore 
bosons due to the kinetic energy, which enjoys as much free
space around each boson as possible.
 
 From Eq.~\eqref{correlation function no superlattice} we deduce
 that the disorder potential created by the frozen bosons has
 slowly decaying correlations. Nonetheless it has fast oscillations
 on short distances, which are captured by the structure factor,
 namely the Fourier transform of the correlator:
 \begin{equation}
\label{correlation function fourier}
C_k=\frac{1}{L}\sum_{r=0}^{L-1}C_r \cos\kla{kr}.
\end{equation}
 This function can be evaluated to give for $k\in[-\pi,\pi]$ and $n^{\rm f}\le 0.5$
 \begin{equation}
 C_k=\begin{cases}
 \frac{1}{L\bigl(1-n^{\rm f}\bigr)}\cdot\frac{|k|}{2k_{\rm F}}&\mbox{for } 
 |k|\le 2k_{\rm F}\\
 \frac{1}{L\bigl(1-n^{\rm f}\bigr)}&\mbox{for } 2k_{\rm F} \le |k|\le \pi.\\
 \end{cases}
 \end{equation}
 The expression for $n^{\rm f}>0.5$ follows from particle-hole symmetry.
 The corresponding functions for an uncorrelated random potential 
generated by randomly displayed frozen particles are  
$C_r=\delta_{r0}$ and $C_k=1/L$. 
We observe that the Fourier spectrum of the frozen-boson
potential is extremely broad, and it has a flat maximum  
for $k \geq k_{\rm F}$,
reflecting the short wavelength oscillations of the potential.
Hence we are in presence of a correlated but strongly fluctuating  
potential. In the following section we will see that such a potential
generally leads to Anderson localization of all single-particle
states and to the suppression of quasi-condensation at all fillings.
Throughout the rest of the paper, we will
take the case of half-filling for the frozen bosons, $N^\text{f}=L/2$.

\section{Ground-State Properties}
\label{Ground State Properties}

In this section we investigate the properties 
of the eigenstates of the Hamiltonian 
${\cal H}_0+{\cal H}_{int}$ (see equations \eqref{H0} and \eqref{Hint}) 
for the mobile bosons moving in the potential created by the frozen 
bosons. Here we consider an homogeneous system
($V=V^\text{f}=0$) with periodic boundary conditions for both species 
of bosons. We start with an analysis of the localization properties in 
real space and continue with an analysis of coherence in momentum space.

We investigate the localization properties in the
frozen-boson potential through the \emph{participation ratio} (PR) 
defined as
\begin{equation}
\label{participation ratio}
{\rm PR}\kla{\ket{\Phi}}=
\frac{\kla{\sum_{i=1}^L\langle n_i\rangle}^2}{\sum_{i=1}^L\langle n_i\rangle^2} ,
\end{equation}
where $\langle n_i\rangle$ is the average particle density on site $i$.
In the case of a rectangular density profile the participation ratio gives 
the support of the density profile, for exponentially localized states it 
is proportional to the decay length of the wavefunction, and for a 
Gaussian-shaped density profile it is proportional to its standard deviation.

\begin{figure}[tb]
\begin{center}
\includegraphics[width=\columnwidth,angle=0]{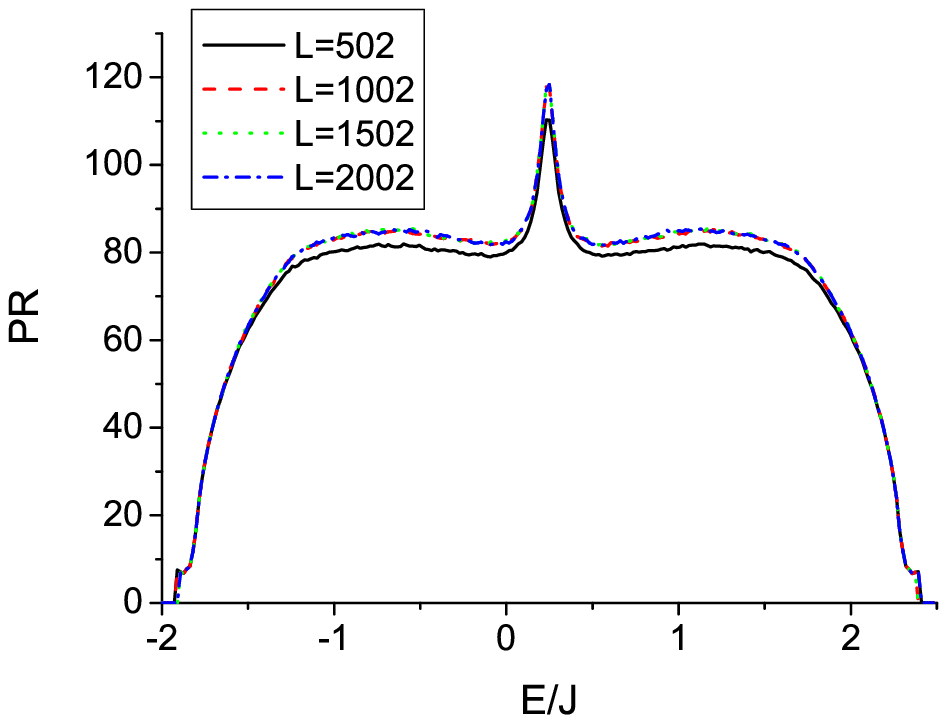}
\includegraphics[width=\columnwidth,angle=0]{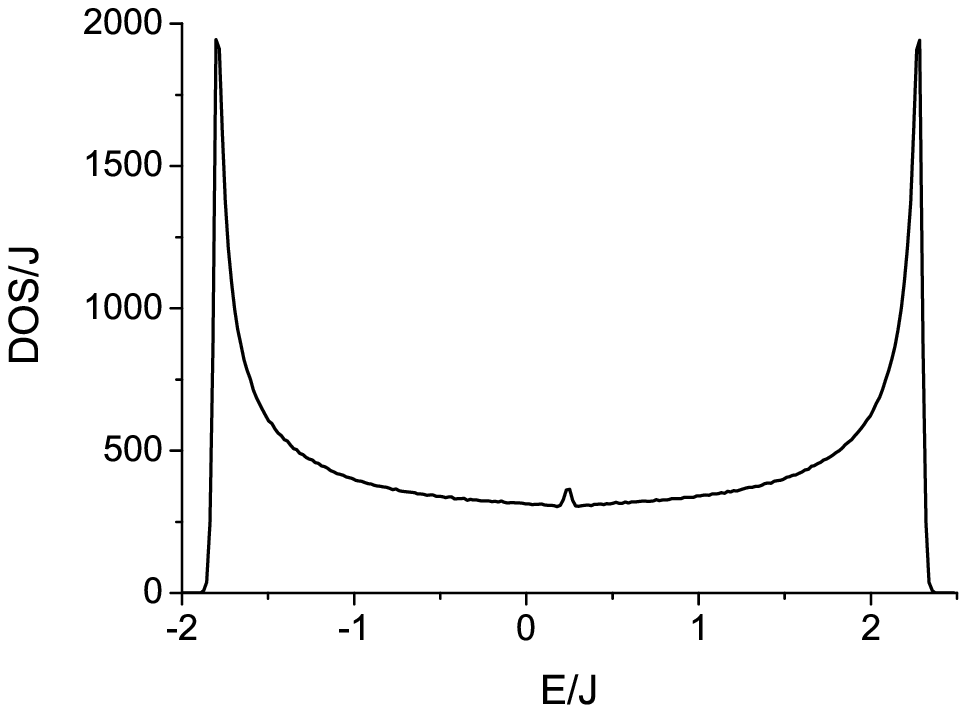}
\caption{(Color online) Average participation ratio (\emph{upper plot})
and density of states (\emph{lower plot}) of the single-particle energy 
eigenstates in the correlated random potential created by frozen bosons with disorder strength $W=0.5J$. The density of
states is calculated for a system size $L=2002$.}
\label{PR eigenstates}
\end{center}
\end{figure}

 The average PR of single particle eigenstates in the potential created
 by a half-filled system of frozen bosons at $W=0.5J$ is shown 
 in the upper panel of Fig.~\ref{PR eigenstates} as a function of 
 energy, for various system sizes. The PR essentially becomes size independent for $L\ge 1000$, clearly indicating the 
 localization of all single-particle
 eigenstates, except for possibly a subset of zero measure;
 we find a similar result for all strengths of the potential
 we investigated. 
 Still a non-extensive number of extended eigenstates, not captured
 by this analysis, is in principle sufficient to suppress 
 localization \emph{e.g.} in transport experiments \cite{Dunlapetal90}, 
 so that further analysis is required to complete the picture on
 the localization  properties of the frozen-boson potential 
 (see Sec.~\ref{Transport Properties}). The peak in the PR at the center of the band can be understood by the fact that the corresponding states have dominant $k=\pm\pi/2$ components, which are also the dominant Fourier components of the random potential for half filling. \\

  We now move on to discuss the properties of the \emph{many-body} 
ground state for hardcore bosons in the frozen-boson potential.
Making contact with the discussion of Sec.~\ref{Hamiltonian Dynamics}, 
such a ground state 
can be reached by adiabatically turning on the interaction $W$
between the two species after having prepared each of them separately
and having quenched the hopping of the frozen bosons.
In particular we analyze the properties of the disorder-averaged
OPDM, which is translationally invariant, so that its eigenvalues
correspond to the momentum distribution function (MDF), and hence average condensation properties 
are studied in momentum space through the scaling
of $\lambda_0=n_{k=0}$ with the number of particles $N$.

\begin{figure}[tb]
\begin{center}
\includegraphics[width=\columnwidth,angle=0]{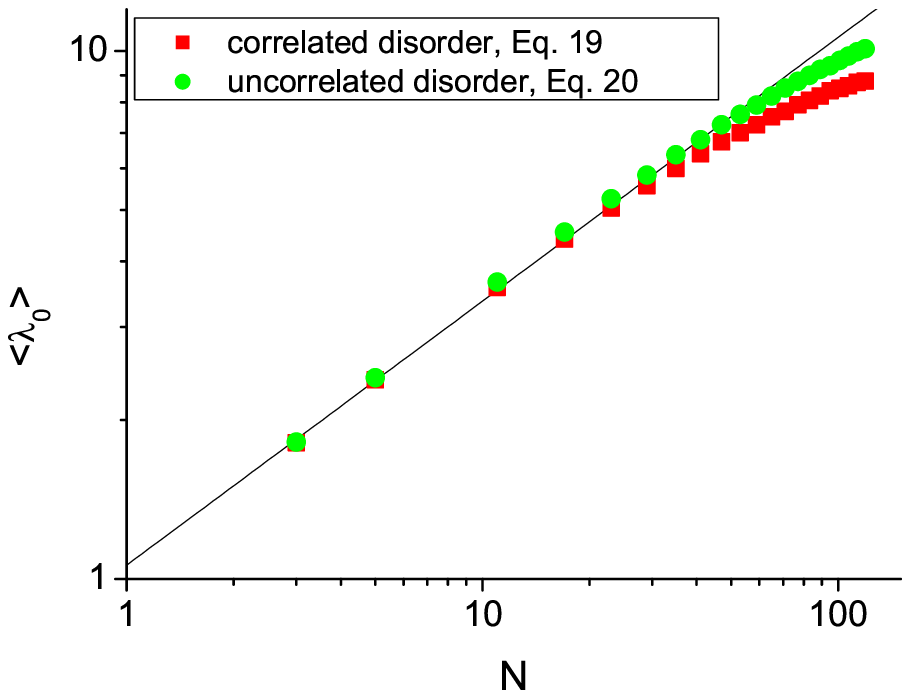}
\includegraphics[width=\columnwidth,angle=0]{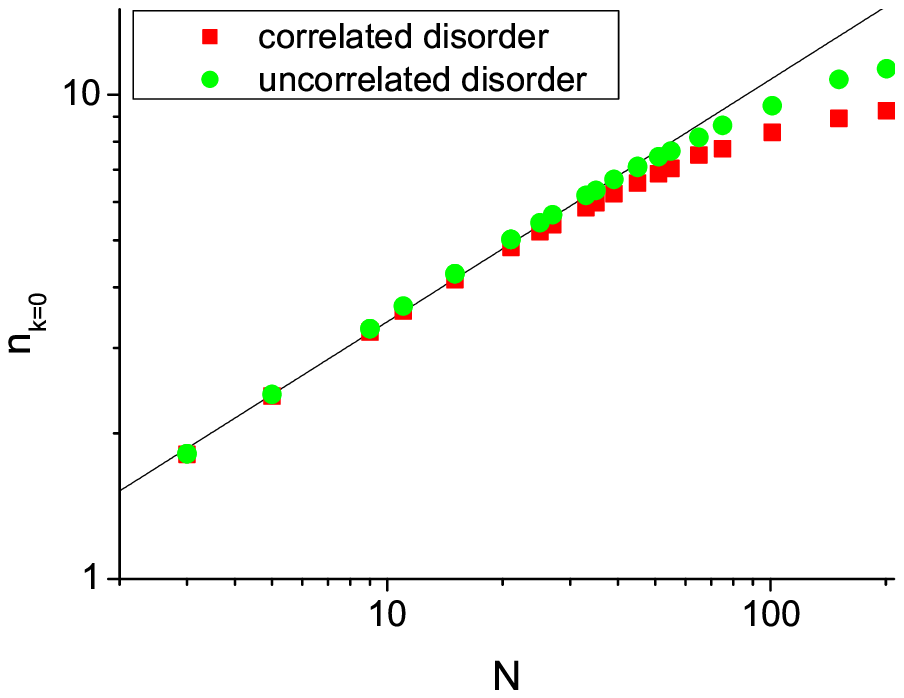}
\caption{(Color online) Scaling of the disorder-averaged largest eigenvalue 
$\lambda_0$ of the OPDM (upper plot) and the occupation of the zero momentum state $n_{k=0}$ (lower plot) for the ground state of a half-filled system of
mobile bosons, interacting with the disorder potential. The cases of a correlated potential
resulting from the superfluid state, Eq.~\eqref{cnif correlated},
and that of a fully uncorrelated potential, Eq.~\eqref{cnif uncorrelated},
are compared.
Here the strength of disorder is $W=0.5J$. 
The black lines are fits to $n_{k=0},\langle\lambda_0\rangle\propto \sqrt N$ 
for the first four data points.}
\label{scaling analysis ground state}
\end{center}
\end{figure}

The scaling analysis of the disorder-averaged occupations of the lowest natural 
orbitals $\lambda_0$ is shown in Fig.~\ref{scaling analysis ground state}
where, as $N$ grows, $N^{\rm f}$ and $L$ are grown correspondingly such that
$N= N^{\rm f} = L/2$.
For small system sizes the hardcore bosons show quasi-conden\-sation behaviour 
$\lambda_0\sim N^\alpha$, with $\alpha=0.5$ within the error given by the 
simulation, but for larger system sizes $\lambda_0$ saturates, hence 
revealing the absence of quasi-condensation in the thermodynamic limit. 
As shown in Fig.~\ref{scaling analysis ground state}, the same quantity 
for the potential generated by fully uncorrelated frozen 
particles gives a completely analogous picture. This crossover can be 
qualitatively explained as a fragmentation effect. The lowest natural 
orbital in each disorder 
configuration is localized, namely it does not scale with the system size
and hence it can can only accommodate a finite number of particles, 
since the bosons repel each other. 

As translational invariance is restored after disorder-averaging, 
the eigenvalues of the disorder-averaged OPDM correspond to the 
momentum distribution function. The peak at zero momentum, in which 
quasi-condensation appears in the case without disorder, is 
significantly reduced in presence of disorder.  
 A scaling analysis of $n_{k=0}$ of 
the mobile bosons, shown in Fig.~\ref{scaling analysis ground state} 
detects a similar crossover from algebraic 
increase to saturation as for the disorder-averaged $\lambda_0$. 
Therefore, the consequences of fragmentation extends to this 
experimentally accessible observable, and we can conclude on the 
absence of quasi-condensation in the disorder-averaged OPDM 
of the ground state of the system \cite{footnote}.
Moreover the density of states, shown in the lower panel of
Fig.~\ref{PR eigenstates}, reveals a continuous excitation 
spectrum at all energies. Hence, for a system of interacting bosons, the absence of quasi-condensation,
together with the absence of gaps in the density of states, leads to classify the state of the system as a \emph{Bose glass}. At the same time the Jordan-Wigner transformation translates this phase into an ideal Anderson insulator of non-interacting fermions.

\section{Dynamical Properties after sudden on-turn of the disorder potential}
\label{Single Trap}
While the previous section investigated the state of the mobile
bosons after an adiabatic on-turn of the interaction with the 
disorder potential created by the frozen bosons, in this 
section we consider the evolution of the mobile bosons after 
turning on that interaction suddenly.
We consider such a time evolution in the homogeneous system, 
\emph{i.e.} $V=V^\text{f}=0$, with periodic boundary conditions,
mimicking the situation in which both species of particles are prepared
in the same region of space before being brought into interaction. 
In this situation, the disorder-averaged real-space density of the mobile
bosons remains uniform during time evolution, so that real-space localization
effects are not visible. Nonetheless, even in a more realistic experimental 
scenario, in which both species of particles are kept in the same 
nonvanishing trap ($V=V^\text{f}>0$), 
the disorder-averaged density profile of the mobile species 
does not reveal marked localization effects, and it is
even expanding during time evolution in order to reduce
the overlap with the confined frozen bosons \cite{Paredesetal05,
Horstmann07}.

\begin{figure}[t]
\begin{center}
\includegraphics[width=\columnwidth,angle=0]{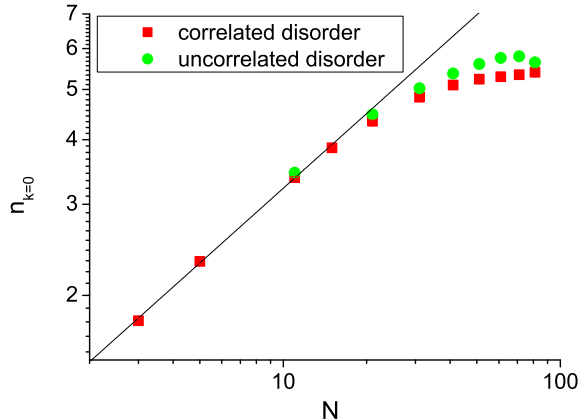}
\caption{(Color online) Scaling of the steady-state zero-momentum
occupation $n_{k=0}$ with the particle number  
$N$ in a homogeneous system of mobile and frozen bosons
($V=V^\text{f}=0$) at half-filling ($N=N^f=L/2$), with interaction strength 
$W=0.5J$. The line corresponds to a fit $n_{k=0}\sim\sqrt{N}$ to the 
first four data points.}
\label{single.MDFscaling}
\end{center}
\end{figure}

 The fundamental effect of disorder on the time evolution 
can be very clearly detected in momentum space, which can be measured 
in time-of-flight experiments. The initial quasi-condensation
peak in the MDF at $k=0$ decreases quickly during time evolution until
a stationary regime is reached, in which the value of $n_{k=0}$ is 
oscillating with a small amplitude ($\sim$ 1\%). 
The scaling analysis for the value of $n_{k=0}$, time-averaged over 
the small oscillations, is shown in Fig.~\ref{single.MDFscaling} for
a half-filled system of  mobile bosons. 
The occupation of the zero momentum state $n_{k=0}$ increases 
approximately like $\sqrt{N}$ for small $N$, but it then deviates from an algebraic 
increase and saturates for larger $N$; a completely analogous behavior
is found for the case of uncorrelated disorder. Hence the scaling of 
$n_{k=0}$ reveals a crossover from quasi-condensation to fragmentation, 
with close similarity to the case of adiabatic evolution
described in Sec.~\ref{Ground State Properties}; in analogy to that
case, we conclude that the steady state reached by time evolution 
realizes \emph{dynamically} a Bose glass.

\section{Transport Properties}
\label{Transport Properties}

 In this section we discuss the expansion properties of initially
confined mobile bosons in the potential created by the frozen particles.
The initial confinement for the mobile particles is provided by
a tight parabolic trap ($V>0$), while the frozen particles
can be imagined as prepared in a much shallower trap, whose
effect is ignored for simplicity, so that the random potential
they generate is the same as the one studied in the homogeneous
case in Sec.~\ref{Characteristics of Disorder}. In particular
we imagine that at time $t=0$ the trap confining the 
mobile bosons is released ($V=0$) and simultaneously 
the two species are brought into interaction ($W>0$).
After release from the trap, the mobile particles would expand
forever in absence of disorder, so that a halt in presence
of disorder explicitly shows localization effects.

Subsection \ref{transport.localization} deals with the evolution 
of the real-space properties during expansion, while Subsection 
\ref{transport.Quasi-Condensation} studies the evolution of 
the condensation properties of the system.

\subsection{Real-space localization}
\label{transport.localization}
\begin{figure}[tb]
\begin{center}
\includegraphics[width=\columnwidth,angle=0]{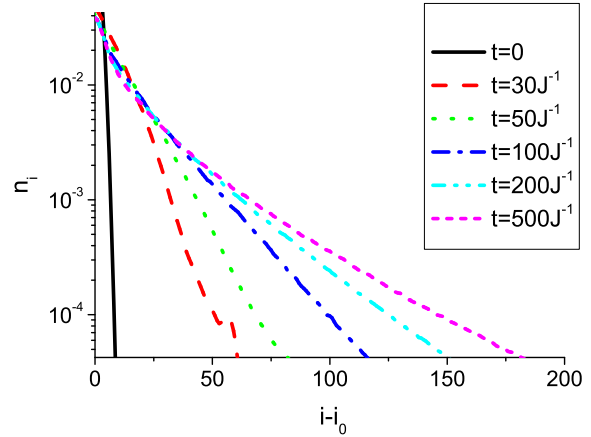}
\caption{(Color online) Snapshot of the evolution of the real-space density $n_i$ 
for a single particle, initially in the ground 
state of a harmonic trap with $V=0.01J$, and interacting with a correlated disorder potential. Here the interaction strength is $W=0.5J$, and the system size is
   $L=502$.}
\label{transport.densities}
\end{center}
\end{figure}

\begin{figure}[tb]
\begin{center}
\includegraphics[width=\columnwidth,angle=0]{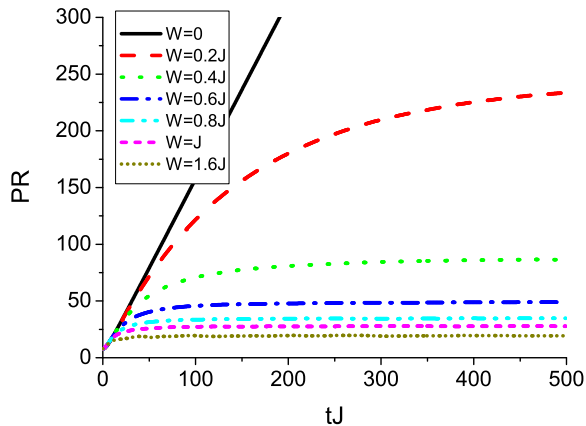}
\caption{(Color online) Time evolution of the participation ratio 
of a single particle for various disorder strengths.
Other parameters as in Fig.~\ref{transport.densities}.}
\label{transport.PR frozen}
\end{center}
\end{figure}

In this subsection we focus on the real-space properties of the 
time-evolved hardcore bosons
in the frozen-boson potential, starting from a confined state in a parabolic trap.
 Snapshots from a disorder-averaged time evolution of a \emph{single} 
 particle starting from the ground state in the trap are depicted in 
 Fig.~\ref{transport.densities}. It is shown that the expansion 
 reaches a \emph{localized} steady state in the long-time limit;
 in particular the spatial decay of the particle density undergoes
 a dramatic change from \emph{Gaussian} - as expected in a parabolic trap - to
 \emph{exponential} - as expected in presence of Anderson localization.
 Hence an Anderson localized state is realized dynamically during expansion;
 it is fundamental to stress that this steady state does not correspond to 
 the ground state of the system, as the energy of the particle
 is conserved during expansion and hence it does not relax
 to the ground-state value.

\begin{figure}[tb]
\begin{center}
\includegraphics[width=\columnwidth,angle=0]{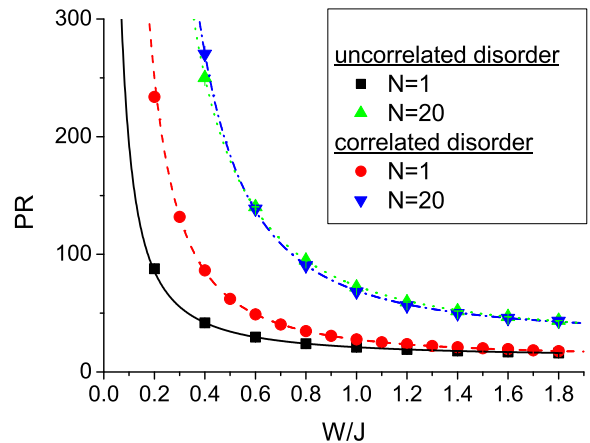}
\caption{(Color online) Steady state participation ratio 
(recorded at $t=500J^{-1}$) for $N$ particles, time-evolved in a system of 
size $L=702$ starting from the ground state of a harmonic trap ($V=0.01J$) 
and interacting with a correlated random potential 
or with a fully uncorrelated random potential. 
The lines show a fit to the data points with an algebraically
decaying function.}
\label{transportPRvarU}
\end{center}
\end{figure}

 The dynamical localization of the single particle wavefunction 
 is fully captured by the time evolution of the PR, depicted
 in Fig.~\ref{transport.PR frozen} for various strengths $W$ of the disorder 
 potential. Without disorder, the wavefunction spreads ballistically without changing
 its Gaussian shape \cite{Rigoletal04}.
 In the presence of disorder, the PR saturates instead to a 
 finite value, which decreases when increasing the disorder potential; 
 a fit to the final value of the PR, shown in Fig.~\ref{transportPRvarU}, 
 suggests that saturation in the PR takes place for any arbitrarily small value 
 of the potential, as it would be expected for uncorrelated
 disorder, although exploring very small strengths of the disorder is 
 numerically demanding as the steady state is reached for 
 prohibitively large values of the PR. 
 
\begin{figure}[t]
\begin{center}
\includegraphics[width=\columnwidth,angle=0]{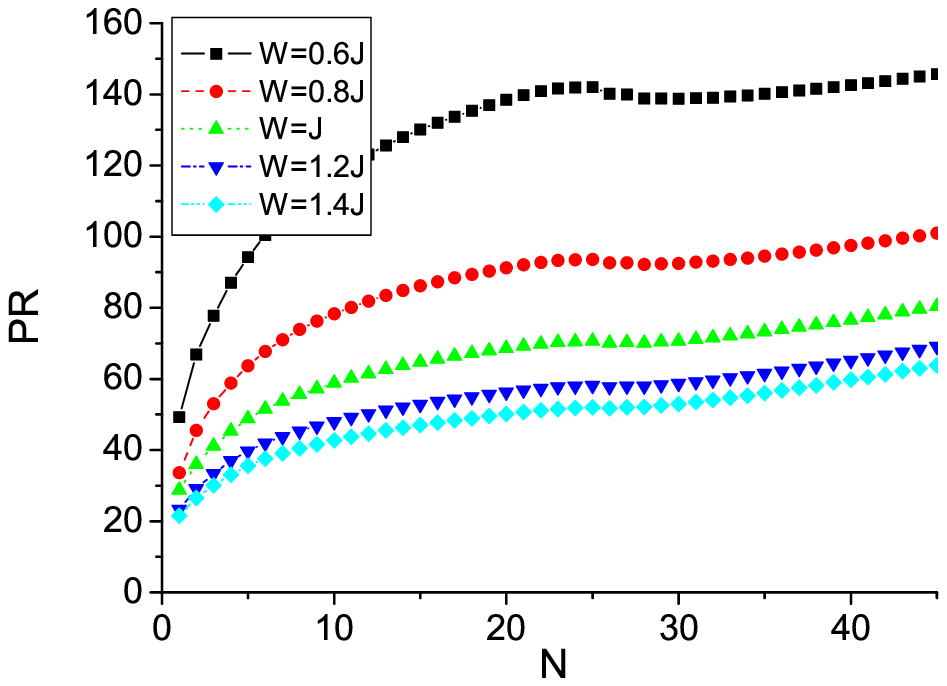} \\
\null\hspace*{.4cm}\includegraphics[width=\columnwidth,angle=0]{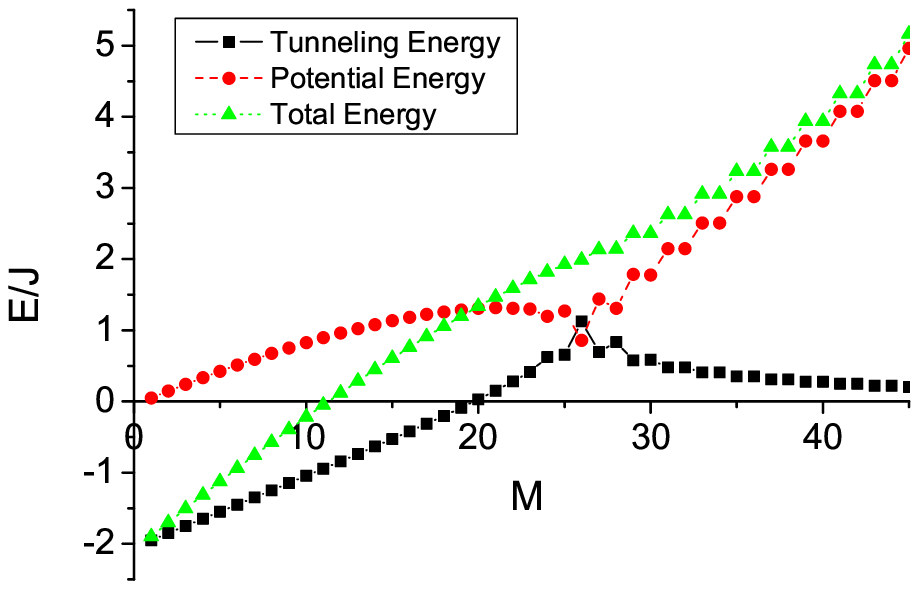}
\caption{(Color online) Upper plot: Steady-state participation ratio 
(recorded at $t=400J^{-1}$)
of $N$ particles as a function of the particle number
for various disorder strengths. Other parameters as in
Fig.~\ref{transportPRvarU}. Lower plot: Total, kinetic, 
and potential 
energy of the $M$th single-particle eigenstate of the initial system 
in the trap.}
\label{transport.varN}
\end{center}
\end{figure}

The time evolution of the participation ratio of many particles initially confined in
a harmonic trap is similar 
to that exhibited by a single particle,
and a saturation of the PR to a steady-state value is observed for the 
smallest value of $W$ that we could treat numerically 
(Fig.~\ref{transportPRvarU}).  \\

Nonetheless, upon increasing the particle number in the trap beyond a critical
value $N_c\approx 26$ (see equation \ref{particle density trap} with $V=0.01$) a Mott plateau at unit filling starts forming in the center, and the  
properties of the initial trapped ground state change drastically. It is 
then interesting to study whether the different initial conditions for the 
time evolution reflect themselves in the steady state of the system after the
halt of the expansion. The upper plot of Fig.~\ref{transport.varN} shows the PR 
of the steady state as a function of the number of particles $N$.
The steady state PR increases with the number of particles for $N\le N_c$,
showing a small peak for $N=N_c$ beyond which the increase with $N$ is
much slower. This feature can be understood in terms of the
time evolution of 
the fermionic wavefunction Eq.~\eqref{matrix representation}, whose real-space 
properties such as the PR are equivalent to those of the hardcore bosons,  
and whose time evolution is simply obtained through the time
evolution of single-particle eigenfunctions. For $N>N_c$ the 
single particle wavefunctions that are populated for increasing $N$
are more and more confined to the sides of the trap, as 
the center has already a saturated density, and their energy 
is dominated by the trapping term, while the single-particle
kinetic energy \emph{decreases} with increasing $N$ beyond $N_c$
(see lower plot of Fig.~\ref{transport.varN}).
Given that the trapping energy vanishes at $t=0$ after the
trap release, the fermions occupying levels beyond the
$N_c$-th expand with an energy which is less than that
of the $N_c$-th level, so that they are expected to be
localized by the disorder potential with a final participation 
ratio which is similar (or even less than) that of the $N_c$-th
particle. The time-evolved particle density of the 
many-body system is the simple sum of the squared time-evolved 
wavefunctions of the free fermions, so that the total participation
ratio is not expected to increase significantly when adding
particles beyond the $N_c$-th.

\subsection{Coherence properties}
\label{transport.Quasi-Condensation}
\begin{figure}[tb]
\begin{center}
\includegraphics[width=\columnwidth,angle=0]{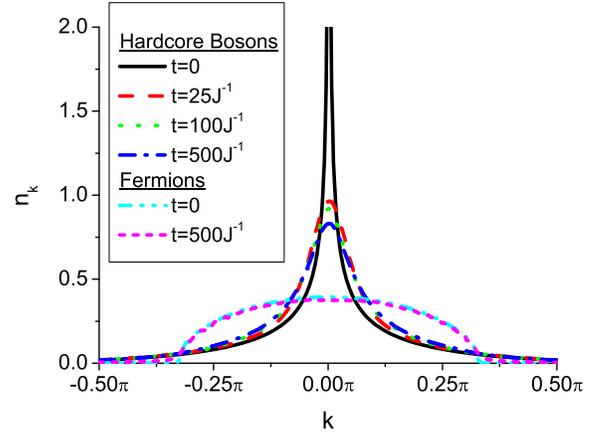}
\caption{(Color online) Snapshots of the time evolution of the 
momentum distribution $n_k$ for a system
of $N=100$ hardcore bosons, initially in a trap with density $\tilde\rho=0.5$, 
interacting with a correlated disorder potential. 
Here the system size is $L=1002$ and the interaction strength is 
$W=0.5J$. The momentum distribution for the corresponding spinless fermions
is shown for comparison.}
\label{transport.trapMDFU05}
\end{center}
\end{figure}
In the previous section the real-space properties of initially confined hardcore
bosons expanding in a disorder potential have been discussed. This
section addresses the condensation properties of the same system, motivated
 by the rich physical scenario offered by expanding hardcore bosons in absence
 of disorder \cite{Rigoletal04,Rigoletal05}. Subsection \ref{transport.Harmonic} is
 devoted to the time evolution of hardcore bosons initially confined in a 
 superfluid state, while Subsection \ref{transport.Mott} analyses the expansion
 from a Mott insulator.

\subsubsection{Time evolution starting from a superfluid state}
\label{transport.Harmonic}
This subsection focuses on the expansion of the hardcore bosons initially
confined in a shallow trap with characteristic
 density $\tilde\rho=0.5$ (see Eq.~\eqref{particle density trap}) well below
 the critical value $\tilde\rho_c\approx 2.6$ for the onset of a Mott plateau
in the trap center. With this initial condition quasi-condensation in the first NO
 is present, and the MDF is peaked around $k=0$ at $t=0$ \cite{Rigoletal05}. 
 In the absence of disorder the MDF evolves towards that of the
 fermions (which is a constant of motion), namely the 
 bosons \emph{fermionize} in momentum space, although they
 still quasi-condense in the lowest NO which contains many
 momentum components \cite{Rigoletal05}.
 Furthermore, the OPDM is decaying algebraically like $\rho_{i,i+r}\sim|r|^{-0.5}$,
 showing quasi-long-range order, whereas the phase of the OPDM is oscillating at
 large distances leading to fermionization in momentum space.  
 The decay of the OPDM is measured from the center of the trap, i.e. the
 center of the lowest NO and the site with maximum occupation.  

\begin{figure}[tb]
\begin{center}
\includegraphics[width=\columnwidth,angle=0]{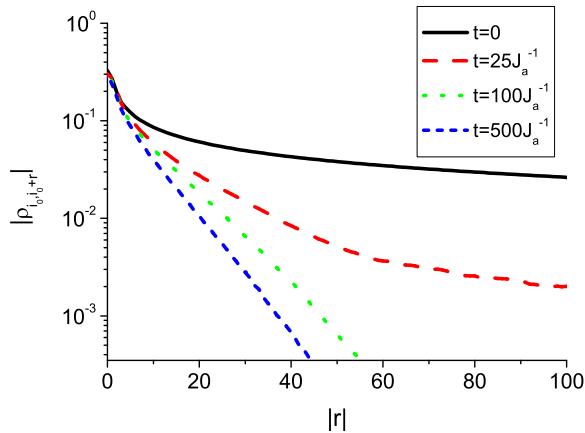}
\caption{(Color online) Decay of the modulus of the OPDM $|\rho_{i_0,i_0+r}|$
from the system center $i_0$ at various times during evolution. 
Parameters as in Fig.~\ref{transport.trapMDFU05}.}
\label{transport.trapOPDMU05}
\end{center}
\end{figure}

We now move on to the analysis of the evolution of the MDF during expansion from
a shallow trap in the presence of disorder created by the frozen bosons. 
The MDF of the mobile hardcore bosons is depicted in Fig.~\ref{transport.trapMDFU05} 
at different times $t$, compared to the MDF of fermions in the same system. 
In presence of disorder the MDF of the fermions is no longer a constant 
of motion, but it is only slightly broadening
in time due to its interaction with the frozen species of particles.
The effect of the disorder potential on the hardcore bosons is more significant.
The $t=0$ peak at zero momentum reduces its height, but, contrary to
the expansion without disorder, it does not disappear 
and fermionization is not present. 
Following Ref. \onlinecite{Rigoletal05}, fermionization in 
absence of disorder is understood through the argument that,
after a long-term expansion, the hardcore-boson system  
is dilute enough to be considered
 essentially non-interacting, and so it becomes equivalent
to its fermionic counterpart. In presence of disorder, on the contrary,
the expansion is stopped by localization, so that the extremely
dilute limit is never reached and the hardcore
bosons preserve their nature of strongly interacting particles. 
Furthermore, the interaction with a disorder potential leads 
to the loss of quasi-long-range order during time evolution, 
as shown in Fig.~\ref{transport.trapOPDMU05} 
by the decay of the OPDM at various times. 
The system finally reaches a steady state with an exponentially 
decaying OPDM. 

\begin{figure}[tb]
\begin{center}
\includegraphics[width=\columnwidth,angle=0]{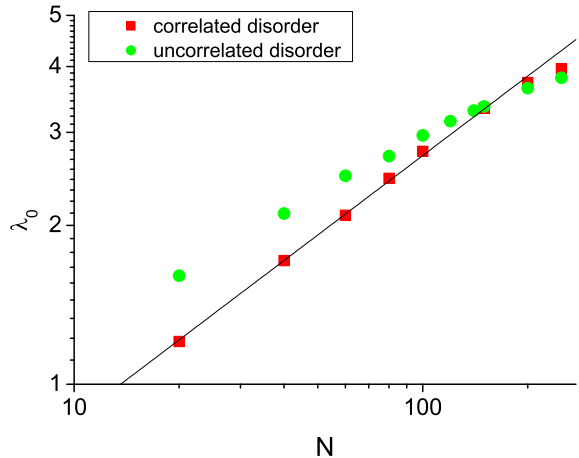}
\caption{(Color online) Scaling of the steady-state $\lambda_0$ 
(recorded at time $t=500J^{-1}$) for varying number of particles in a system of size $L=1502$;
other parameters as in Fig.~\ref{transport.trapMDFU05}. 
The data for correlated frozen bosons are compared with those
for fully uncorrelated ones. 
The line corresponds to a fit to $\lambda_0\sim \sqrt{N}$.}
\label{transport.trapscaling}
\end{center}
\end{figure}  

The loss of quasi-long-range order during expansion strongly
suggests the loss of quasi-condensation in the system.
As the initial conditions break the translational symmetry, condensation properties 
are not captured by the scaling of the occupation of the zero-momentum 
state,
and direct diagonalization of the disorder-averaged OPDM is necessary to extract
the scaling of the occupation $\lambda_0$ of the lowest natural orbital. 
$\lambda_0$ decreases during 
time evolution, reaching a constant value in correspondence with the steady
state observed in real-space. The scaling analysis of $\lambda_0$ is performed 
for this constant value in Fig.~\ref{transport.trapscaling}. 
The scaling of $\lambda_0$ deviates from the quasi-condensation behavior
at sufficiently large particle numbers, revealing a crossover from 
quasi-condensation to fragmentation due to the localized nature
of the lowest NO. 
 Fully uncorrelated disorder leads to a similar behavior, although 
 the crossover appears to be much broader, 
and a quasi-condensation regime at low $N$ could not be identified.

\subsubsection{Time Evolution starting from a Mott insulator}
\label{transport.Mott}

\begin{figure}[tb]
\begin{center}
\includegraphics[width=80mm,height=50mm,angle=0]{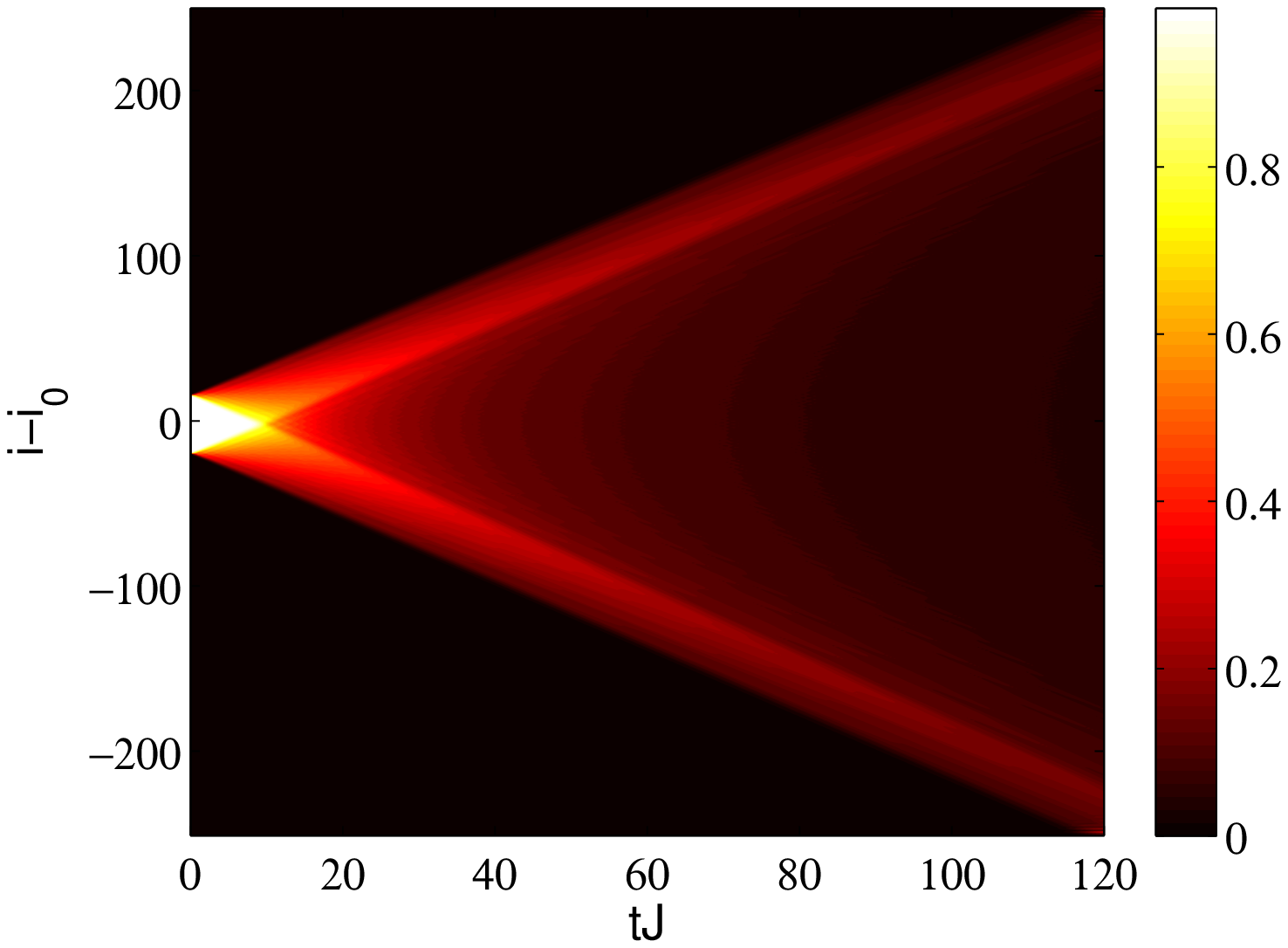}
\includegraphics[width=80mm,height=50mm,angle=0]{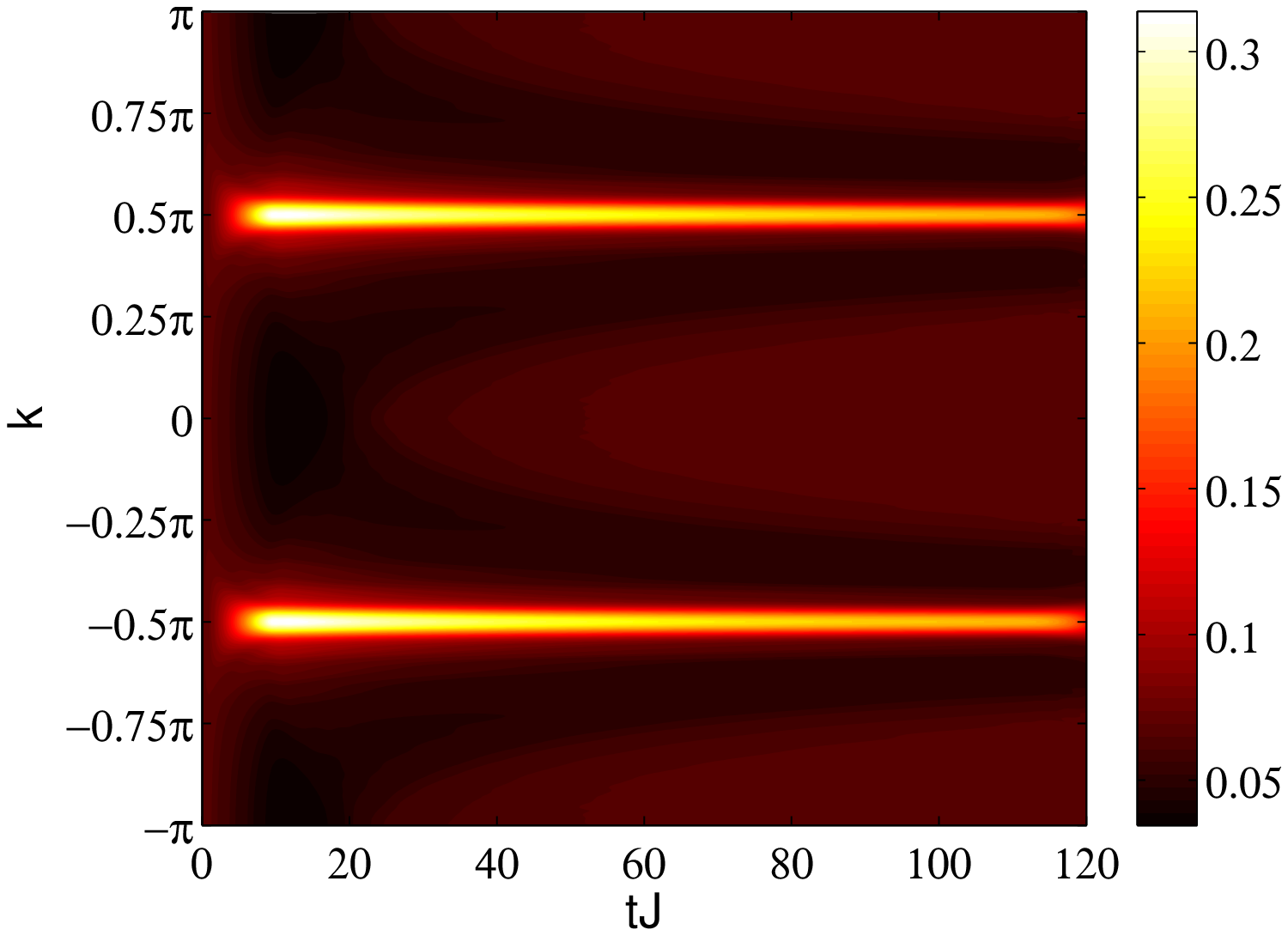}
\caption{(Color online) Time evolution of the density profile (upper image) 
and the momentum distribution function (lower image) of $N=35$ hardcore bosons, 
initially in a perfect Mott insulator, in a system of size $L=502$ without disorder.}
\label{transport.MottU0}
\end{center}
\end{figure}

\begin{figure}[tb]
\begin{center}
\includegraphics[width=80mm,height=50mm,angle=0]{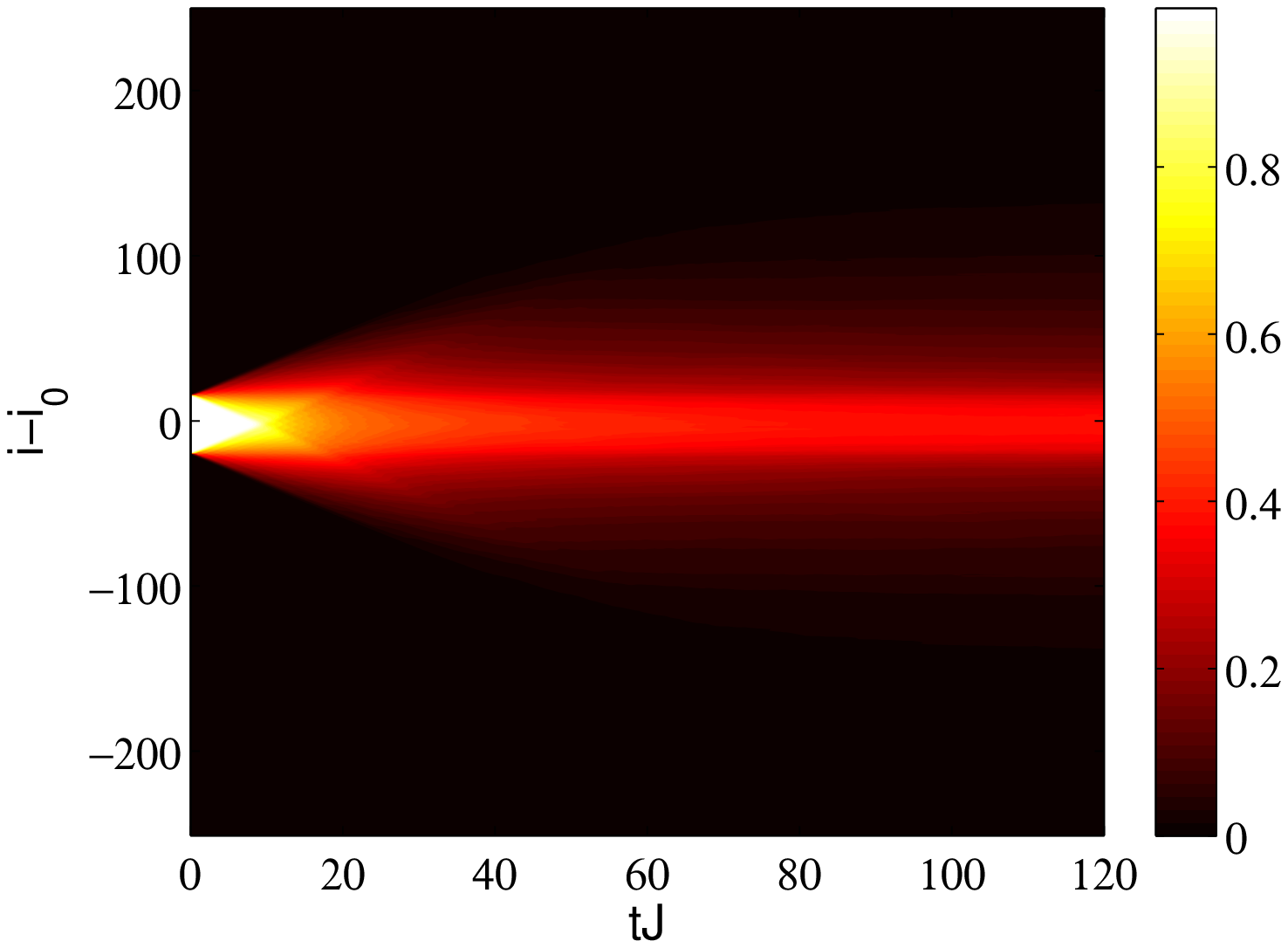}
\includegraphics[width=80mm,height=50mm,angle=0]{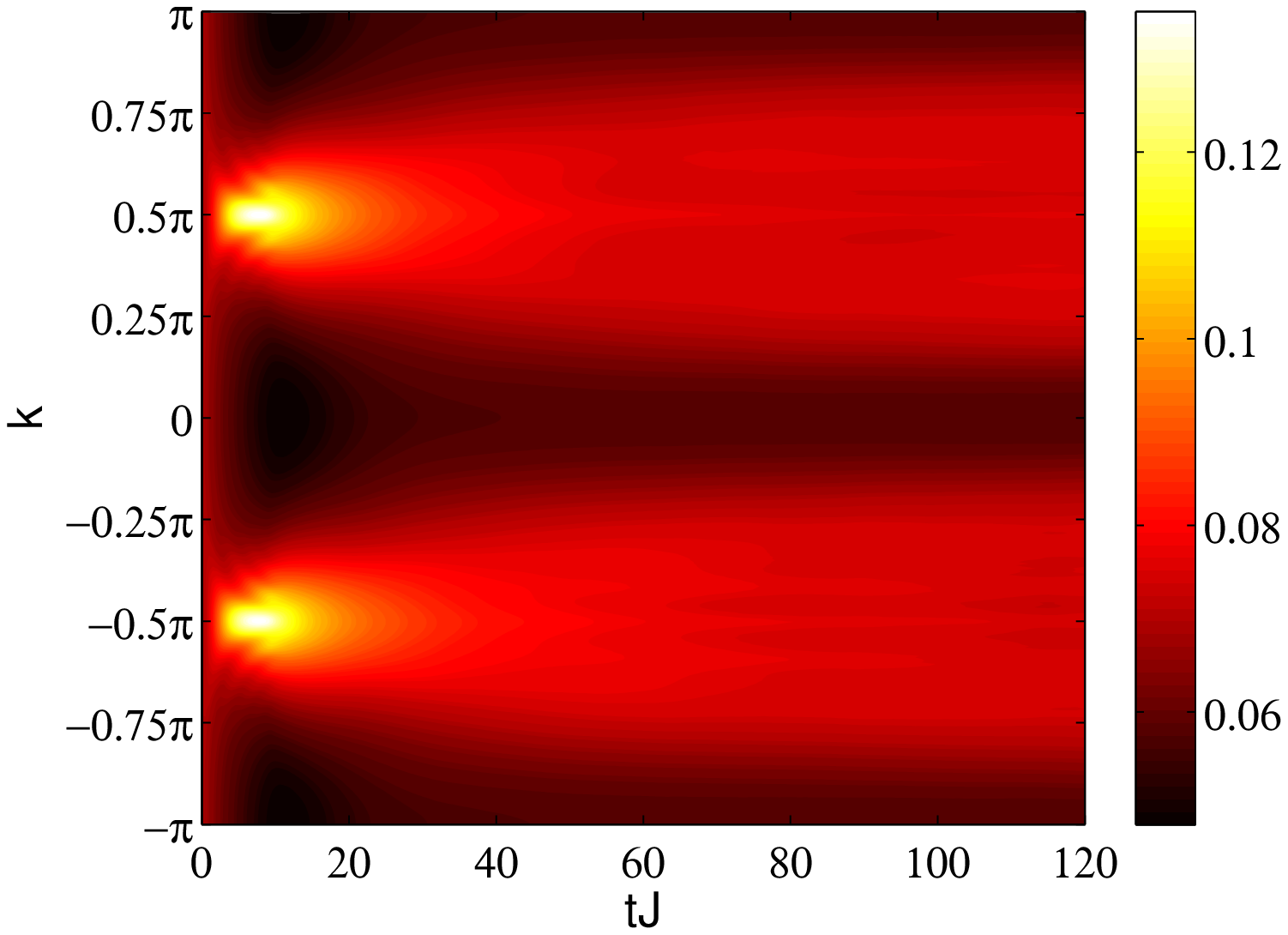}
\caption{(Color online) Time evolution of the density profile (upper image) 
and the momentum distribution function (lower image) of $N=35$ hardcore bosons, 
initially in a Mott insulator, interacting 
with the disorder potential created by frozen particles at half-filling.
Here the interaction strength is $W=0.5J$.}
\label{transport.MottU05}
\end{center}
\end{figure}

As reported in Ref.~\onlinecite{Rigoletal04}, hardcore bosons prepared
in a perfect Mott insulating state (corresponding to an infinitely
steep trap) and subsequently time evolved exhibit the phenomenon 
of dynamical quasi-condensation at finite momentum. Indeed,
from the initially flat MDF of the Mott insulator two quasi-condensation
peaks emerge at momenta $k=\pm\pi/2$ during time evolution \cite{Rigoletal04}, 
as reproduced in Fig.~\ref{transport.MottU0}. Strictly speaking 
quasi-condensation happens
in two degenerate NOs whose Fourier transform is sharply peaked around 
$k=\pm\pi/2$; the NOs propagate at a velocity $v=\pm 2J$ 
corresponding to the maximal group velocities 
$\partial\epsilon_k/\partial k$ for the single-particle dispersion relation 
$\epsilon_k=-2J\cos k$ at momenta $k=\pm\pi/2$. 

The occupation of the degenerate lowest NOs, $\lambda_0$, follows
initially a universal power-law increase with time, independent
of the number of particles $N$; at an $N$-dependent 
characteristic time $\tau_c$, the two degenerate lowest NOs begin to move in opposite directions, and 
$\lambda_0$ starts to algebraically decrease, although
its scaling with particle number shows the typical 
quasi-condensation behavior $\lambda_0 \sim \sqrt{N}$.
The two lowest NOs clearly 
appear in the real-space densities as coherent fronts of the atomic cloud
moving in opposite directions 
(see Fig.~\ref{transport.MottU0}). This aspect suggests the use of this
setup to produce an atom laser \cite{Rigoletal04}. 

We now consider the case of expanding hardcore bosons in the disorder 
potential created by frozen particles. Fig.~\ref{transport.MottU05}
shows the density profile and the MDF 
for the expansion from a perfect Mott insulator in the presence of 
disorder created by frozen particles. Initially peaks 
at momenta $k=\pm\pi/2$ are emerging from the flat MDF at $t=0$ 
as in the case without disorder, and the hardcore boson cloud shows 
two outer fronts expanding ballistically in opposite directions. Nonetheless
this initially coherent expansion is rapidly suppressed due to
localization, which leads to a decrease and broadening of the 
momentum peaks at larger times up to a final steady-state
MDF in which two broad peaks survive,
and which corresponds to a localized state in real space as seen
in Subsection \ref{transport.localization}.

\begin{figure}[tb]
\begin{center}
\includegraphics[width=\columnwidth,angle=0]{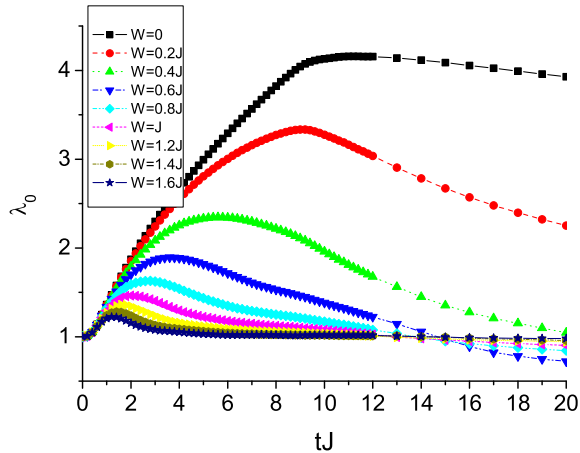}
\caption{(Color online) Time evolution of $\lambda_0$ for various 
disorder strengths. Other parameters
as in Fig.~\ref{transport.MottU05}.}
\label{transport.lambdavarU}
\end{center}
\end{figure}

\begin{figure}[tb]
\begin{center}
\includegraphics[width=\columnwidth,angle=0]{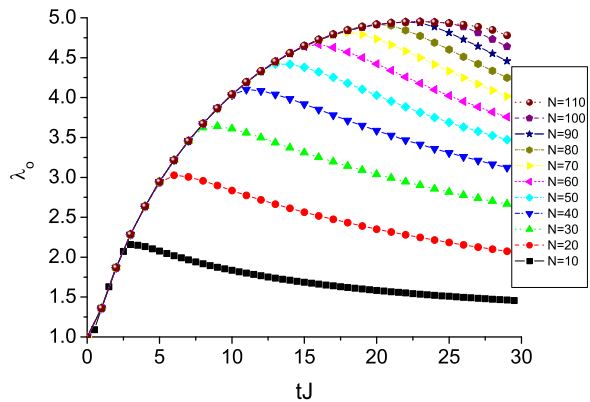}
\caption{(Color online) Time evolution of $\lambda_0$ for various
numbers of particles, and for disorder strength $W=0.1J$. Other parameters
as in Fig.~\ref{transport.MottU05}.}
\label{transport.lambdavarN}
\end{center}
\end{figure}

A deeper analysis of condensation effects relies on the NOs and their 
occupations. Figs.~\ref{transport.lambdavarU} and \ref{transport.lambdavarN} 
show the time evolution of the largest eigenvalue $\lambda_0$ for different 
disorder strengths and different particle numbers. Initially $\lambda_0$
is two-fold degenerate, 
corresponding to reflection symmetry at the center of the system,
and it increases following a universal power law independent of 
the particle number, similarly to what is observed in absence
of disorder but with a different disorder-dependent exponent. 
As in the case $W=0$, at a characteristic time $\tau_c$ the 
two degenerate natural orbitals start to move into opposite 
directions, and correspondingly the time evolution of 
$\lambda_0$ turns into a decreasing behavior; $\tau_c$ strongly depends
on the disorder strength (Fig.~\ref{transport.lambdavarU}), 
while its dependence on the particle number $N$ becomes
weaker for large $N$, where $\tau_c$ is seen to approach
an asymptotic value (Fig.~\ref{transport.lambdavarN}).
Unlike the case $W=0$, at a later stage the expansion
of the system is stopped by disorder, and the 
degeneracy in the lowest NOs is removed, going
from two propagating ones to a single NO localized 
in the system center.

\begin{figure}[tb]
\begin{center}
\includegraphics[width=\columnwidth,angle=0]{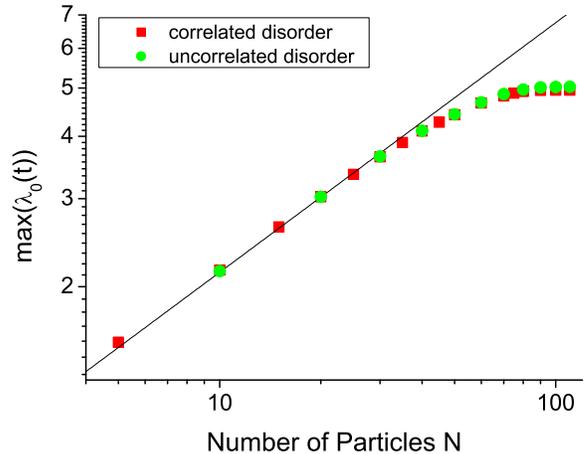}
\caption{(Color online) Scaling of the maximal value of $\lambda_0$ during time 
evolution of $N$ hardcore bosons, initially in a Mott insulator. All parameters
as in Fig.~\ref{transport.lambdavarN} . The line corresponds to a fit 
to $\lambda_0\sim\sqrt{N}$ of the 
first four data points.}
\label{transport.Mottscaling}
\end{center}
\end{figure}

 Hence the expanding system of hardcore bosons from a pure
Mott state shows a crossover from an incipient quasi-condensation
regime at finite momenta to a localization regime. It is natural
to ask whether the system displays true quasi-condensation at any
intermediate point in time. To address this issue, we perform
a scaling analysis of $\lambda_0$ 
at the maximum-coherence time $t= \tau_c$.
The results of this analysis are shown in Fig.~\ref{transport.Mottscaling}, 
where $\lambda_0$ exhibits a clear saturation for large particle
numbers, and hence a crossover from quasi-condensation to 
fragmentation. Repeating the same analysis for uncorrelated random disorder 
we find a scaling of $\lambda_0$ which is in qualitative agreement 
with the case of disorder generated by frozen particles (the apparent
quantitative agreement is accidental).

\section{Experimental Realization}
\label{Experimental Realization}

\begin{figure}[h!!]
\centering
\includegraphics[width=\columnwidth]{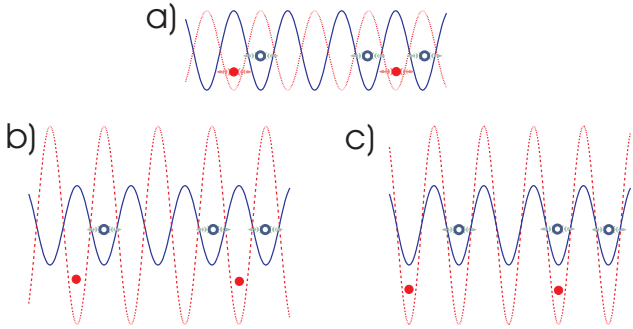}
\caption{(Color online) Sketch of the preparation
of the disorder potential and of the tuning of its strength 
using state-dependent optical lattices for a mixture
of atoms in two internal states. The two species
are first prepared in two shifted optical lattices
\emph{(a)}, then one of the two optical lattice
is increased in strength so as to freeze one of
the two bosonic species \emph{(b)}, and finally the
spatial phase of the two lattices is changed to 
bring the two species into interaction \emph{(c)}.}
\label{two species interaction}
\end{figure}

 The original experimental proposal motivating this work 
is reported in Ref.~\onlinecite{Paredesetal05}, where the
use of state-dependent lattices was initially envisioned
to create the quantum superposition of random potentials.
As shown in Ref.~\onlinecite{Mandeletal03}, two hyperfine
states of the same atomic species can be loaded in the
minima of two different polarization components of 
an optical lattice at a so called "magic wavelength"
\cite{Jakschetal99,Brennenetal99,toolbox},
at which each hyperfine state couples to one and only one 
polarization component. Hence, if the two polarization 
components are initially shifted by $\pi/2$ in space,
the two species are non essentially interacting 
(see Fig.~\ref{two species interaction}). The possibility
of increasing drastically the intensity of one of the
two polarization components would allow for a sudden quench 
of the hopping of one of the two species, preparing in this
way the quantum superposition of random potentials. 
The two species can be then brought into interaction
at different strengths by shifting the spatial phase
between the two polarization components of the lattice.
An adiabatic shift would transfer the mobile species
to the Bose-glass ground state in the random potential, as discussed
in Section \ref{Ground State Properties}, whereas a sudden 
shift would give rise to an out-of-equilibrium Bose-glass 
state as steady state after a transient evolution, as
discussed in Section \ref{Single Trap}. As demonstrated
in Ref.~\onlinecite{Paredesetal04}, the hardcore regime  
is achieved by using
deep lattices for both polarization components and 
extremely dilute gases of both hyperfine states, 
so that a density of less than one atom per site is
achieved in the lattice after loading. 

 As seen in Section \ref{Transport Properties}, expansion experiments
in the disorder potential require to confine the two 
components with strongly different trapping frequencies,
and then to release one of the two traps independently
of the other. Making use of the selective coupling of
different hyperfine states to different polarization
components of a magic-wavelength laser, it is possible
to more strongly confine one of the two species through
an optical dipole trap obtained by a tightly focused
and polarized laser propagating transverse to the chain 
direction in the one-dimensional optical lattice. 

 Throughout all the previous Sections we have seen
that the onset of localization leaves very strong signatures
on the momentum distribution, which is observed in standard
time-of-flight experiments. Moreover expansion experiments
in the disorder potential lead to exponentially localized steady
states, whose density profile can be reconstructed by 
absorption images taken shortly after turning off all lasers
\cite{Lyeetal05, Clementetal05, Fortetal05, 
Schulteetal05}. Despite the averaging over the various
tubes of the one-dimensional optical lattice \cite{Paredesetal04}, 
the extreme tails of the averaged particle density distribution 
should be dominated by the exponentially localized tails of 
the particle density in the central tubes of the one-dimensional 
optical lattice.

\section{Conclusions}

 In this paper we have shown that a gas of one-dimensional 
hardcore bosons undergoes genuine quantum localization effects
when set into interaction with a secondary species of 
bosons frozen in a massive quantum superposition of Fock states.
Each Fock state can be regarded as a realization of a
random potential, and the unitary evolution of the 
mobile species of bosons follows all possible paths
related to the various disorder realizations \emph{in parallel}.
Physically relevant states in which one can prepare the
frozen bosons, as for instance the superfluid state in 
the hardcore limit, realize a rapidly fluctuating disorder 
potential over the length scale of a few lattice spacings;
despite its power-law decaying correlations, the disorder 
potential is found to lead to the same localization 
effects as those observed in a \emph{fully uncorrelated} potential.   
In the hardcore boson limit for the mobile species these 
effects can be studied exactly in real time through 
Jordan-Wigner diagonalization, and we can numerically
simulate realistic experimental setups for the dynamical
preparation of localized many-body states. 
The equilibrium state of the hardcore bosons in the 
random potential is found to be a homogeneous Bose-glass 
state with exponentially decaying correlations; a similar
state can be realized also dynamically after a sudden
on-turn of the interaction between the two species. 
When the hardcore bosons are initially confined in a 
tight trap and then set free to expand in the random 
potential, for any non-vanishing
disorder strengths the expansion stops and the system
reaches an exponentially localized state.

 For any setup discussed in this work, we observe the absence of 
quasi-condensation and quasi-long-range order due to the disorder potential. 
When the two species are confined in the same region and 
brought into interaction, the steady state of the system displays
exponentially decaying off-diagonal correlations;
in the expansion experiments the disorder potential
destroys the effects of fermionization (when expanding from an initially
dilute state) and quasi-condensation at finite momentum (when expanding
from a Mott state). 
Hence disorder created by a species of frozen hardcore bosons represents 
a very robust way to experimentally implement strongly fluctuating random 
potentials in optical lattices, and to realize fundamental localization
effects of many-body systems. 

\section{Acknowledgements}

 We thank J.~J. Garc\'ia-Ripoll and V. Murg for useful discussions. This work
 is supported by the European Union through the SCALA integrated Project.


\begin{thebibliography}{99}
\bibitem{Sanchezetal07} Nevertheless, recent theoretical works has
shown that laser-speckle potentials within experimental reach
can possibly lead to Anderson localization of weakly interacting
bosons \cite{Kuhnetal05,Sanchezetal07b,Clementetal06,Luganetal07}.
\bibitem{footnote} Although the disorder-averaged largest eigenvalue 
$\langle \lambda_0 \rangle$ and the disorder-averaged $k=0$ occupation
$n_{k=0}$ show a similar scaling with system size, they have no simple 
relationship to each other; indeed $\lambda_0$ refers to localized 
natural orbitals, while $n_{k=0}$ is associated with an extended one.
\bibitem{remarkuncorrelated} The uncorrelated frozen bosons are considered to be in the state $\ket{\Phi^{\rm f}}=\otimes_{i=1}^L(1+a_i^\dagger)\ket{0}$ projected on the subspace with a fixed number of particles. This state cannot straightforwardly be prepared in an optical lattice.
\end{thebibliography}
\end{document}